\documentclass[aps,amsmath,amssymb,prl,twocolumn,showpacs,superscriptaddress]{revtex4}

\usepackage{amsfonts}
\usepackage{graphicx}

\newcommand{\rvec}{\mathbf{r}}

\newcommand{\Rvec}{\mathbf{R}}
\newcommand{\Kvec}{\mathbf{K}}

\newcommand{\avec}[1]{\mathbf{#1}}
\newcommand{\dee}{\mathrm{d}}

\newcommand{\qvec}{\mathbf{q}}

\begin{document}

\title{Splitting of roton minimum in the $\nu=5/2$ Moore-Read state}

\author{Anthony R. Wright}
\affiliation{Institut f\"ur Theoretische Physik, Universit\"at Leipzig, D-04103, Leipzig, Germany}
\affiliation{Max-Planck-Institut f\"ur Festk\"orperforschung, D-70569 Stuttgart, Germany}
\author{Bernd Rosenow}
\affiliation{Institut f\"ur Theoretische Physik, Universit\"at Leipzig, D-04103, Leipzig, Germany}
\affiliation{Max-Planck-Institut f\"ur Festk\"orperforschung, D-70569 Stuttgart, Germany}

\begin{abstract}
We calculate the dynamical structure factor of the $\nu=5/2$ non-abelian Moore-Read state in the dipole approximation, valid for large momenta. Due to the fact that both quasi-particles (qps) and quasi-holes (qhs) have an internal Majorana degree of freedom, a qp-qh pair has a fermionic degree of freedom which can be either empty or occupied, and leads to a splitting of the roton mode.  Observation of this splitting by means of finite wavelength optical spectroscopy could provide evidence for Majorana modes in the $\nu=5/2$ quantum Hall state. 

\end{abstract}


\maketitle

The $\nu=5/2$ quantum Hall (QH) state is expected to support non-abelian quasi-particles (qps), which could be used  for topological quantum computation \cite{simonrev}. There is both theoretical and experimental evidence that the $\nu=5/2$ QH state is indeed described by the non-abelian Moore-Read (MR) state \cite{MR91}. Most significantly, the quasiparticles in the MR state have charge $e/4$. This has been confirmed by tunneling \cite{radu}, shot noise \cite{dolev} and interference \cite{willett} experiments. However, photoluminescence experiments seem to indicate that the state is not spin polarized, contrary to the MR state \cite{mstern}. Recent resonant light scattering experiments suggest that both polarized and unpolarized domains may form in the second Landau level in general \cite{rhone}. 
Recent experiments employing nuclear magnetic resonance find evidence for a spin polarized ground state \cite{Tiemann+12,Stern+12}. 
Clearly more evidence is needed to confirm the validity of the MR state (or its particle-hole conjugate \cite{apf}, which for the purpose of this letter is identical)

The roton mode is the collective excitation of quantum Hall states \cite{kh,mpg}. In integer quantum Hall systems, it can be understood as a particle hole pair, in fractional quantum Hall (FQH) systems as a quasi-particle (qp) quasi-hole (qh) pair. For hierarchical FQH states, there is one roton minimum for each level of the hierarchy. 
The roton mode can be probed with the help of optical excitations, but only in the presence of finite wavelength 
density modulations. Using surface acoustic waves to create a charge density modulation, several roton minima for hierarchical FQH states were recently observed experimentally \cite{smet}.

The non-abelian statistics of $\nu=5/2$ MR qps is encoded by an internal  Majorana  degree of freedom. 
Here we show that as a direct consequence of these  Majorana fermions, the roton mode in the $\nu = 5/2$ MR state, in the presence of disorder-pinned qps, splits into two branches. 
The two Majorana modes of a qp-qh pair can be combined into a neutral fermion, which
can be either occupied or unoccupied.  Due to the finite overlap of the two Majorana wave functions associated with the qp and qh, an occupied neutral fermion has a different energy than an unoccupied one, hence the splitting of the roton mode.  The splitting both oscillates and decays exponentially with increasing separation of the pair.  The main result of this work is the excitation spectrum given by Eq.~(\ref{dispn}), whose form is shown in Fig.~1. We argue that the  splitting of the roton mode and hence the possibly non-abelian nature of the $\nu = 5/2$ state
can be directly probed  in an optical experiment similar to \cite{smet}.

\begin{figure}[tbp]
\centering\includegraphics[width=8cm]{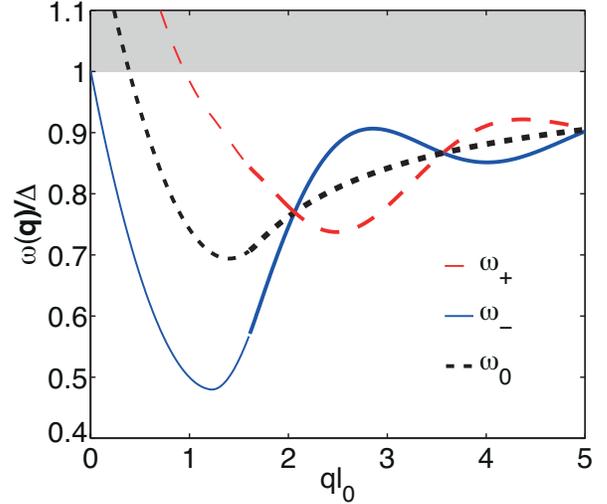}
\caption{(Color online) The expected roton dispersion. The thick lines are given by Eq.~(\ref{dispn}), constituting the dipole approximation,  valid for $q$ beyond the roton minimum, which occurs near $ql_0 \approx 1.4$ \cite{jain52}. The usual single roton mode (shown with black dots) has split into two branches with distinct extrema, denoted $\omega_+$ and $\omega_-$, where $\omega_+-\omega_- = 2E(2l_0^2q)$. $\omega_-$ is the brighter primary branch, and $\omega_+$ is the secondary split-off branch. In the the small $q$-regime, the dipole Hamiltonian is no longer valid and the roton minimum is taken to be parabolic with appropriate parameters used as outlined in the text. This regime is plotted with thin lines. $\Delta \approx 0.025e^2/\epsilon l_0$ is the $\nu=5/2$ exciton gap \cite{morf2}. The grey shaded region denotes the two-roton continuum.}
\label{dispnfig}
\end{figure}

We model the MR state as a p-wave superconductor of composite fermions \cite{rg}, and then use
a Ginzburg-Landau approach to describe the dynamics of the superconducting order parameter.
 Charge $e/4$
qps of the MR state are vortices of the superconductor, and are treated in the same way as 
the vortices of condensed composite bosons \cite{zhk,zhk2} in order to calculate the roton branch. The dynamics of the Ginzburg-Landau order parameter effective theory with $N$ vortices are identical to those of an $N$ point particle Hamiltonian with pairwise interactions and a constant kinetic term per particle, with the commutation relations $[\hat{x}_i,\hat{y}_j] = -il_0^2\delta_{i,j}/\nu$, where $i$ denotes the $i^\mathrm{th}$ vortex, and $l_0 = \sqrt{\hbar/eB}$ is the magnetic length.

Vortex configurations of the order parameter in the Bogolubov-de-Gennes (BdG) Hamiltonian give rise to low-energy Majorana excitations, pairs of which form neutral fermions. Our Hilbert space then includes both the vortex degrees of freedom as well as the occupancy of neutral fermions that reside in pairs of vortices. We shall truncate the many-body Hilbert space of vortices to a single vortex-antivortex pair, precisely the qp-qh pair that forms the roton mode.

For Majoranas separated by $r$, the unoccupied and occupied energies of the neutral fermion in a qp-pair have been calculated for a p-wave superconductor \cite{ds}. Using the same approach to calculate the splitting for a qp-qh pair, and using the qh wavefunction from Ref. \cite{simpwave}, we obtain for $r>l_0$
\begin{equation}
	E(r) = \mp\frac{2\Delta}{\pi^{3/2}}\frac{\cos(p_F r + \pi/8)}{\sqrt{p_Fr}}\mathrm{exp}\biggl(-\frac{r}{\xi}\biggr),
	\label{split}
\end{equation}
in which $\Delta$ is the mean field value of the order parameter field, $p_F$ is the Fermi velocity, and $\xi$ is the Majorana coherence length, which is valid for $r>l_0$. To make quantitative predictions for the MR state, we use the estimates of  Ref.~\cite{bar}
for  the  magnitude of the splitting near the roton minimum ($\approx 0.01e^2/\epsilon l_0$),  the  coherence length, $\xi\approx 2.3l_0$, and for $p_F \approx \pi/3l_0$.

The Hamiltonian for a single qp-qh pair in this effective picture is
%
\begin{equation}
	\hat{H}_0(\hat{r}) = \Delta + V(\hat{r}) + E(\hat{r})(2\hat{c}^\dag \hat{c} - 1)
	\label{H0}
\end{equation}
%
Where $\Delta$ is the energy gap to create two isolated qps \cite{morf2}, $\hat{r} = |\hat{\rvec}_1 - \hat{\rvec}_2|$ is the dipole separation operator, $V$ is the Coulomb interaction between the charge $\pm e/4$ vortices, and $\hat{c}^\dag = \hat{\gamma}_1 - i\hat{\gamma}_2$ is the neutral fermion creation operator which contains two Majorana operators ($\hat{\gamma}_1$ and $\hat{\gamma}_2$). The non-vanishing commutators  in relative separation ($\hat{\mathbf r}$) and centre of mass ($\hat{\mathbf R}$) coordinates are $[\hat{r}_x,\hat{R}_y] = 2il_0^2$, and $[\hat{r}_y,\hat{R}_x] = -2il_0^2$. The canonical centre of mass momentum is then related to the relative separation as $\hat{\rvec} = 2l_0^2\hat{z}\times\hat{\Kvec}$, where $\hat{z}$ is the out-of-pane unit vector. The Hamiltonian has the dynamics of a dipole drifting with constant velocity in a magnetic field, with an internal 2-level degree of freedom.

For qp-qh separations comparable to the intrinsic size of a vortex, the dipole approximation breaks down, and the leading contribution to the dynamic structure factor is of quadrupole type.  For separations much larger than the intrinsic  size of a vortex, the single dipole approximation is asymptotically exact \cite{zhk2}.

Formally, the ground state of the Hamiltonian Eq.~(\ref{H0}) 
 is at $r = 0$, with both a divergent splitting energy Eq.~(\ref{split}) and a divergent potential $V$.  These divergences are an artefact   of the point-particle description of vortices: the Hamiltonian Eq.~(\ref{H0}) is only valid for well separated qps, ie.~$r$ greater than twice the intrinsic size of a single vortex. However, in the following calculation we 
use a completely overlapping qp-qh pair as the vacuum state and assign the energy zero to it. Then, the density operator defined below separates the qp and qh from each other and creates a finite energy excitation described by 
Eq.~(\ref{H0}).

The parity of the number of quasiparticles must be conserved: an excitation breaks a Cooper pair, necessarily incorporating \emph{two} fermions. It is necessary then, to add to the Hamiltonian Eq.~(\ref{H0}) a set of vortices pinned by disorder, which can, in pairs, support neutral fermions. These vortices will inevitably be present in any real system due to disorder, but their density can also be tuned externally.

We are interested in the dynamics of a single qp-qh pair, created by a momentum fluctuation within this background of pinned vortices. For sufficiently large vortex densities a lattice forms in which the mid-gap BdG states become dispersive. In the clean lattice limit the dispersion is gapped. However, due to the oscillatory overlap integral, Eq.~(\ref{split}), it is highly disorder-dependent and regains a large zero-energy density of states as well as a continuum of midgap states if the lattice spacing is allowed to vary \cite{sterntriangle, trebst}. We avoid this problematic regime entirely by demanding low vortex densities. The arrangement of pinned vortices is assumed to be random. If the average separation of two pinned vortices is greater than a few times the qp-qh separation, the overlap of the dipole vortices with the disorder vortices will be appreciable for \emph{at most} one pair of disorder sites. Thus, in the spirit of the random singlet phase \cite{BL81}, we can discard all other pinned vortices, and reduce our system to that of four vortices: the qp-qh pair and the two closest pinned vortices. Let the operators $\hat{c}$ and $\hat{c}^\dag$ correspond to the dipole neutral fermion, and $\hat{b}$ and $\hat{b}^\dag$ to the single pinned vortex pair. We can now introduce a tunneling term to the Hamiltonian,

\begin{equation}
\begin{split}
	\hat{H}_T = t_d \hat{b}\hat{c} + t_d^*\hat{c}^\dag \hat{b}^\dag + t_d' \hat{b}^\dag \hat{c} + t_d'^*\hat{c}^\dag \hat{b}  + (2 \hat{b}^\dagger \hat{b} -1) \epsilon_d
	\label{HT}
\end{split}
\end{equation}
Here,  $t_d$ and $t_d'$ are the complex overlap integrals between the four sites ($1,2,3,4$), and $\epsilon_d$ denotes  the energy splitting of the localized neutral fermion. We obtain $t_d = E(|\rvec_1 - \rvec_3|) - E(|\rvec_2 - \rvec_4|) + i(E(|\rvec_2 - \rvec_3|) + E(|\rvec_1 - \rvec_4|))$, where $t_d'$ has a similar form but is irrelevant here. We can now proceed to calculate the collective excitation spectrum, given by $\omega(q) = \int \omega S(q,\omega)\dee\omega/\int S(q,\omega)\dee\omega$ \cite{mpg}, where the dynamic structure factor is given by

\begin{equation}
\begin{split}
S(q,\omega) &= \int_{-\infty}^\infty \dee t e^{i\omega t}\langle \hat{\rho}_q(t)\hat{\rho}_{-q}(0)\rangle.
\end{split}
\end{equation}
Here $\hat{\rho}_{q}(t)$ is the charge density operator with time dependence $\hat{\rho}_q(t) = e^{i\hat{H}t}\hat{\rho}_q(0) e^{-i\hat{H}t}$, in which $\hat{H} = \hat{H}_0 + \hat{H}_T$. For a single qp-qh pair, $\hat{\rho}_q(0) = \sum_{s=\pm1}se^{i\qvec\cdot(\frac{s}{2}\hat{\rvec} + \hat{\Rvec})}$. We calculate the eigenstates of $H_0$ explicitly, and perturbatively introduce $H_T$ up to second order. Fourier transforming, we obtain two terms, $S_0(q,\omega)$ which contains only the unperturbed Hamiltonian, and $S_1(q,\omega)$ which contains $\hat{H}_T$ to second order. The first term produces the primary excitation branch in which the neutral fermions are unoccupied. That the neutral fermions are unoccupied is immediately obvious as the charge density operator acts only on the vortex sector of the ground state Hamiltonian. The second term contains only those terms $\propto t_d t_d^*$, as the mixed creation-annihilation operator terms vanish. This term in the structure factor contains operators that fill the two neutral fermion states, and produces the secondary split-off branch, peculiar to the $ \nu=5/2$ MR state.

The total Hilbert space is a product of the vortex space with matrix elements of the form $\langle 0|\hat{\rho}_q|\Kvec\rangle$, and the neutral fermion space with matrix elements of the form $\int_0^t dt_1 \int_0^{t_1} dt_2 \langle 0 | \hat{H}_T(t_1)\hat{H}_T(t_2)|m,n\rangle$, where $m$ and $n$ correspond to the occupancy of the disorder-pair and the dipole respectively, and the time dependence of $\hat{H}_T(t)$ arises from using the interaction picture with respect to $\hat{H}_0$. The unusual temporal limits reflect that for times $t_{1(2)}<0$ or $t_{1(2)}>t$, the system is in the vacuum state and so cannot support Majoranas in the dipole, making $\hat{H}_T$ nonzero only in the time interval $[0,t]$. The vortex part of the ground state Hamiltonian is diagonal in relative separation ($\hat{\rvec}$), and the centre of mass coordinates do not appear. The wavefunction can be written as 

\begin{equation}
	\psi_\Kvec(\rvec,\Rvec) = \frac{1}{\sqrt{V}}e^{i\Kvec\cdot\Rvec}\delta(\rvec + 2l_0^2\hat{z}\times\Kvec).
	\label{eq:}
\end{equation}
This is a point particle description of the vortices. However, in order to calculate the overlap of two vortices, the vortices must be given a spatial profile whose width reflects the size of the vortex. We introduce a Gaussian profile, such that $\delta(\rvec + 2l_0^2\hat{z}\times\Kvec)\rightarrow 2l_0/\sqrt{\pi}e^{ -|\rvec + 2l_0^2\hat{z}\times \Kvec|^2/4l_0^2}$. For a charge-density fluctuation of momentum $q\hat{x}$, we obtain a single solution which is a qp-qh pair with separation $2l_0^2q\hat{y}$, and matrix element $2\exp(-l_0^2q^2/2)$, thus obtaining

\begin{eqnarray}
S_0(q,\omega)& = &  \int_{-\infty}^\infty \dee t \int \frac{\dee\Kvec}{4\pi^2}\sum_{m,n} e^{-i(\epsilon_{mnK} - \epsilon_{000} - \omega)t} \nonumber \\
& & \cdot |\langle 0,0,0|\hat{\rho}_\avec{q}|m,n,\Kvec\rangle|^2
\end{eqnarray}
where the eigenvalues are
\begin{equation}
	\epsilon_{mnK} = \Delta + V(2l_0^2K) + (2n-1)E(2l_0^2K) + (2m - 1)\epsilon_d
	\label{eq:}
\end{equation}
and $\epsilon_d$ is the contribution from the pinned vortex Majoranas. As the density operator is simply the real space translation operator, the summation in the Majorana sector gives $\delta_{m,0}\delta_{n,0}$. Therefore this part of the structure factor is the primary branch of Eq.~(\ref{Sqw}), without the decreased spectral weighting of $M_Q^2$. As for $S_1(q,t)$, the first order expansion of $H_T$ gives zero, and for the second order term we find
\begin{equation}
\begin{split}
S_1(q,\omega) &=  \frac{t_d t_d^\star}{4} \int \dee t \dee t_1 \dee t_2 \sum_{m,n}\int \frac{\dee^2 K}{4\pi^2}  e^{-i(\epsilon_{mn\Kvec}-\epsilon_{00\Kvec})(t_1-t_2)}\\
&e^{-i(\epsilon_{00\Kvec}-\epsilon_{000} + \omega)t}|\langle 0,0,0| \hat{\rho}_q \hat{c}\hat{b}|m,n,\Kvec\rangle|^2 \ \ .
\end{split}
\end{equation}
Here the summation over the Majorana sector results in a single occupied neutral fermion term $\delta_{m,1}\delta_{n,1}$ due to the operator pair $\hat{c}\hat{b}$ and its complex conjugate. Therefore this term contributes  the delta functions  in Eq.~(\ref{Sqw}).
We thus obtain the structure factor, and therefore excitation spectrum, of our qp-qh pair system where the vortices support neutral fermions with non-zero overlap integrals, on top of a sparse background of qps and qhs.

The final form of the structure factor up to second order in $H_T$ is

\begin{equation}
\begin{split}
	&S(Q,\omega) = \frac{2}{\pi}e^{-Q^2/4l_0^2}\biggl[\bigl(1 - M^2_Q\bigr)\delta(\omega_-(Q) - \omega) \\
	&\,\,+ M^2_Q\bigl(\delta(\omega_+(Q) - \omega) + E(Q)\delta'(\omega_-(Q) - \omega)\bigr)\biggr] \ \ . \\
	\label{Sqw}
\end{split}
\end{equation}
Where $Q = 2l_0^2q$, and $M_Q = |t_d|/4(\epsilon_d+E(Q))$, and $\delta'$ denotes the derivative of the delta function with respect to its total argument. This latter term effectively doubles the matrix element of the secondary branch, such that the spectral weighting of the higher energy branch is $2M_Q^2$, and the lower branch has spectral weighting $1 - 2M_Q^2$. Solving for the excitation spectrum, we obtain two distinct branches given by

\begin{equation}
\omega_\pm(Q) = \Delta + V(Q) \pm E(Q) 
	\label{dispn}
\end{equation}
Here, the splitting $\epsilon_d$ of the localized fermion has been neglected because in the relevant parameter regime it will be much smaller than the $E(Q)$ due to the relatively large distance between localized vortices. However, we do include $\epsilon_d$ in our estimate of $M_Q$, as discussed in the supplemental material.

The splitting of the two excitation branches is a direct result of vortices supporting Majoranas, and the exponentially decaying splitting demonstrates that well separated Majoranas have approximately zero energy. These are two necessary ingredients for the vortices to be non-abelian quasiparticles. The expected dispersion is shown in Fig.~1. 
Experimentally, the most important signature is the crossing of the two roton branches near $q \ell_0 =2$. As a consequence, the roton minimum has split into two and shifted to higher and lower $q$ values, respectively. 
The parameters used for $E(Q)$ are outlined below Eq.~(\ref{split}). We have also taken the roton gap to be half the exciton gap $\Delta\approx 0.025e^2/{\epsilon l_0}$, the energy required to create a maximally separated qp-qh pair \cite{morf}, which agrees well with exact diagonalization studies \cite{greiter92, quinn10}, and have set the roton minimum to occur at $ql_0\approx 1.4$, in good agreement with numerics \cite{jain52}.

The two quantities $E(Q)$ and $t_d$ are independently tunable. The former determines the amount of splitting between the two branches and is a function of the separation of the dipole (which is proportional to the momentum at which the roton branch is probed). The latter, $t_d$, determines the spectral weighting of the secondary branch, but will also broaden both branches, and is a function of the proximity of the dipole to the closest pinned vortex pair, and therefore is directly related to the disorder density.

The parameter $t_d$ must be tuned such that the broadening of the excitation spectrum lines does not fuse the two branches. However, $t_d$ must be sufficiently large that a measurably bright higher energy band, whose intensity is proportional to $|t_d|^2$ is observed. In the supplemental material, we present a conservative estimate for the intensity of the secondary branch relative to the primary branch, demonstrating that a relative intensity of $>15\%$ for two clearly resolved peaks is feasible. In addition, we demonstrate that the magnetophonon mode of localized qps is energetically well separated from the roton mode.

In addition to the MR state, alternative wavefunctions have been proposed for the description of the $\nu=5/2$ state. Most notably,  the $331$ candidate wavefunction differs from the MR state in that the spins are unpolarized. We can obtain information about the roton excitation of the 331-state by using the fact that this wavefunction has been used to describe pseudo-spin correlations in $\nu_\mathrm{tot}=1/2$ bilayer QH systems. For bilayer systems, the inter-layer Coulomb interaction takes the form $V(r) \propto (r^2+d^2)^{1/2}$, where $d$ is the interlayer spacing. The relevant case for $\nu = 5/2$ then, is to take the $d\rightarrow 0$ limit of these results. Two magnetoroton spectra have been predicted for $\nu_\mathrm{tot}=1/2$, due to in-phase and out-of-phase pseudo-spin density fluctuations \cite{bls}, and these persist in the $d\rightarrow0$ limit. If the 331 wavefunction correctly describes $\nu=5/2$, then the following observations can be made. The splitting between the two branches is a minimum at the roton minima, contrary to our result. Depending on the splitting magnitude, the upper branch may be lost in the excitation continuum altogether. Furthermore, the splitting will not decrease exponentially (or even at all) with increasing $q$. Finally, the splitting will not oscillate. The combination of these signatures allows for a clear distinction from the splitting of the roton minimum for the Moore-Read state as described above.

Secondly,  a `generalized composite fermion' picture of $\nu=5/2$ was recently discussed and  illustrated by exact diagonalization \cite{quinn10}. In order to understand the lowest energy excitations,  the possibility of two types of elementary excitations was assumed: a magnetoroton consisting of a qp-qh pair, and a broken composite fermion pair. 
This scenario for $\nu=5/2$ was raised already some time ago \cite{greiter92}. From the numerical work \cite{quinn10}, it seems possible that the gap for a broken pair type excitation  may  be slightly below the magnetoroton gap. However, the broken pair spectrum is a Bogolubov spectrum that monotonically increases with increasing $q$, while the magnetoroton spectrum on the other hand asymptotically approaches the exciton gap with increasing momentum. For momenta relevant to our current calculation then, that is for $q$ beyond the roton minimum, the broken pair energy will be larger than the magnetoroton energy, and will disappear into the continuum region.
Most importantly, composite fermions are charge neutral and are expected to couple only weakly to the charge density operator. For this reason, it is unclear whether the broken pair mode can be detected experimentally when studying the structure factor $S(q,\omega)$.

Finally, we comment on the neutral fermion gap: the energy required to add a single unpaired composite fermion to the system \cite{bonderson, jainucf}, which is comparable to the exciton gap $\Delta$ \cite{bonderson}. The neutral fermion dispersion also contains roton-like minima \cite{moller}, however an experimental probe must change the particle number parity \cite{jainucf}. Thus the collective excitation spectrum presented here will not include the neutral fermion dispersion.

We have shown that the magnetoroton dispersion of the $\nu = 5/2$ MR state splits into two distinct branches due to a  fermionic degree of freedom associated with  qp-qh pairs. This neutral fermion consists of a pair of Majorana states in the vortex cores which are responsible for the non-Abelian statistics of the Moore-Read state. In order to overcome the requirement of parity conservation, we coupled the qp-qh pairs to localized qps in the bulk of the quantum Hall state. The intensity of the secondary branch of the roton dispersion is proportional to the square of the coupling strength to these bulk qps. Experimental observation of this splitting by means of optical spectroscopy would constitute compelling evidence for the validity of the Moore-Read state and thus non-Abelian quasiparticles in the $\nu=5/2$ QH state.

Acknowledgements: We would like to thank B.I.~Halperin, S.~Simon, and J.~Smet  for helpful discussions and acknowledge financial support by BMBF.

\section{Supplemental Material for ``Splitting of roton minimum in the $\nu=5/2$ Moore-Read state''}

\subsection{Estimating the intensity of the split-off branch}
The intensity  of the split-off branch in the roton spectrum 
depends on the density of localized quasiparticles. In analogy to less exotic quantum Hall states \cite{puddles}, even in samples which are tuned to the center of the $5/2$ quantum Hall plateau, spatial disorder in the charge distribution in the doping layer can be  expected to give rise to 
 finite density of localized quasi-particles (qps) and quasi-holes (qhs). As the density of these disorder induced   qps and qhs is difficult to estimate, we focus on qps 
 due to a deviation of the actual filling factor $\nu$ from the ideal filling factor $5/2$ in the following. Consider a filling factor of $\nu=5/2 +\eta$, where $\eta$ is small enough such that the system remains on the $\nu=5/2$ plateau ($|\eta| \lesssim 0.01$) \cite{pfaff}. For concreteness, we assume $\eta > 0$, such that there is a finite density of localized qps. 
   
We need to estimate $M_Q^2(\eta) = |t_d(\eta)|^2/16(\epsilon_d(\eta)+E(Q))^2$, which determines the relative intensity of the split-off branch in the expression for the structure factor Eq.~(9) of the main paper. We take the average of the tunnel coupling squared between one qh  which is part of the roton excitation, and one intrinsic qp localized by disorder to be $[|t_d(\eta)^2|]_{av} \approx  [E(r)^2]_{av}$, where $E(r)$ is given Eq.~(1) of the main paper. Here, $[...]_{av}$ denotes the disorder average over distances $r$
between qp and qh.  In order to obtain an upper bound for the energy splitting  
$[\epsilon_d(\eta)]_{av}$ of a  localized
qp-qh pair, we note that the average distance between localized qps is greater than or equal to the average separation between one localized and one externally excited qp and qh, and so $[|\epsilon_d(\eta)|]_{av} \lesssim  [|t_d(\eta)|]_{av}$. To avoid spurious divergences in $M_Q^2(\eta)$ near crossing points between the two roton branches, and using the above estimate for $|\epsilon_d(\eta)|$, we introduce $\tilde{M}_Q^2(\eta)$ as a conservative approximation and lower bound for the  matrix element $M_Q^2$
%
\begin{equation}
\tilde{M}_Q^2(\eta) = \frac{[|t_d(\eta)|^2]_{av}/16}{[|t_d(\eta)|^2]_{av}+E^2(Q)} \lesssim M_Q^2(\eta).  
\end{equation}
%

In order to compute $[|t_d(\eta)|^2]_{av}$, we need to know the average distance $R$ between localized quasiparticles. For a filling fraction $\nu=2.5+\eta$, the excess flux density is $\rho_\phi = \eta/(2\pi l_0^2)$, which gives a quasiparticle density
\begin{equation}
\rho_{qp} = \frac{\eta}{\pi l_0^2}
\end{equation}
due to the deviation from the center of the $5/2$-plateau. In order to simplify the calculation of $[|t_d(\eta)|^2]_{av}$, we associate with each localized qp  a disk with radius $R$ such that $\rho_{qp} =\frac{1}{\pi R^2}$, and hence 
\begin{equation}
R = \frac{l_0}{\sqrt{\eta}} \ \ .
\end{equation}
We now compute  the average tunnel coupling squared by integrating over separations $r$ inside the disk with radius $R$ and find
\begin{equation}
\begin{split}
[t_d(\eta)|^2]_{av} &= \rho_{qp}\int_0^R \! \! \!  r \; \mathrm{d}r  \int_0^{2\pi} \! \! \!    \mathrm{d}\phi \; 
|E(r)|^2\\[.5cm]
& =  \eta \Delta^2 \frac{6 C }{\pi^4} + O(\exp(-2R/\xi)) \ ,  \\
\end{split}
\end{equation}
where $C = 2.3(1 + 1/(\sqrt{2}(1+2.3\pi/3)))\approx2.8$. Putting everything together, we find
\begin{equation}
\begin{split}
\tilde{M}_{q}^2(\eta) \approx \frac{C\eta  ql_0/16}{C\eta ql_0 + \cos^2(2\pi ql_0/3 + \pi/8)e^{-4ql_0/2.3}} \ .
\end{split}
\end{equation}

Next, we discuss the broadening of both the primary and the split-off roton branch due to the tunnel coupling 
to localized qps. We describe the magnitude of the broadening by
\begin{equation}
\gamma(\eta) = \sqrt{\langle |t_d(\eta)|^2\rangle} \approx \sqrt{6C \eta}\Delta/\pi^2 \ \ ,
\end{equation}
and incorporate it in Eq.~(9) (main paper) for the structure factor by replacing
 $\delta(x)\rightarrow \frac{\gamma}{\pi}(x^2+\gamma^2)^{-1}$. Despite this broadening of the delta-function, there is a significant enhancement of intensity near an extremum of the roton dispersion relation $\omega(q)$,  which increases the relative intensity of the split-off branch in the vicinity of its minimum near 
 $q \ell_0 =2.5$. We would like to note that the intensity enhancement due to an increased density of states near  an extremum of the roton dispersion is so strong that the energy of the roton minimum can be determined even in experiments without momentum resolution \cite{dsmomentum,pin} . 

\begin{figure}[tbp]
\centering\includegraphics[width=8cm]{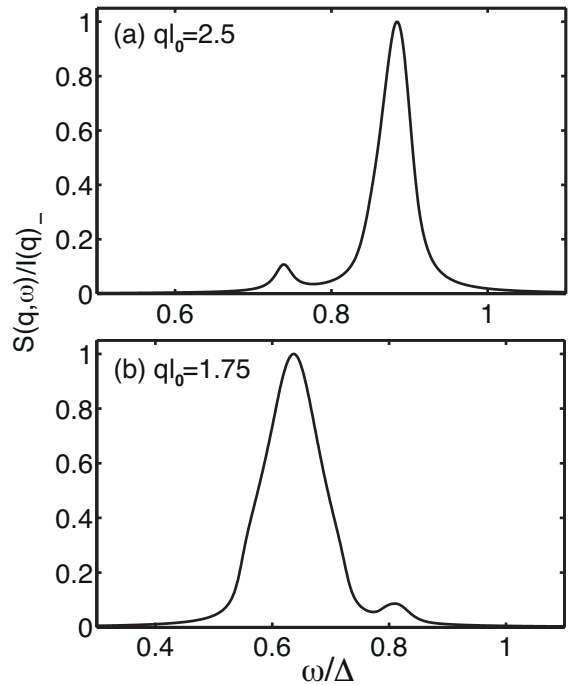}
\caption{The intensity of the dynamic structure factor at filling factor $\nu=2.501$, for two representative momenta, normalized by the primary branch intensity for each plot.  In this experimentally relevant region, the intensity of the split-off line is as much as 15\% of the primary branch and can be clearly resolved. For these plots, the broadening in momentum direction is $\alpha=0.1l_0^{-1}$.}
\label{crosssection}
\end{figure}

Finally, the experimental resolution of features in the structure factor is limited by momentum non-conservation due to scattering off localized qps, which breaks translational symmetry. A recent experimental study of the roton mode in several fractional quantum Hall  \cite{smetp} states allows us to estimate the broadening in momentum direction as $\alpha \approx 0.1l_0^{-1}$. This broadening is 
incorporated into the structure factor by convoluting it with a Lorentzian of width $\alpha$, 
%
\begin{equation} 
{S}_{\alpha,\eta}(Q,\omega) = \int_{-\infty}^{\infty} \mathrm{d}Q' {S}_\eta(Q',\omega)\frac{\alpha/\pi}{(Q-Q')^2+\alpha^2} \ \ , 
\end{equation}
%
where
\begin{equation}
\begin{split}
	&S_{\eta}(Q,\omega) = \frac{2}{\pi}e^{-(Q)^2/4l_0^2}\biggl[(1 - 2 \tilde{M}_{Q}^2(\eta))\frac{\gamma/\pi}{(\omega_-(Q) - \omega)+\gamma^2} \\
	&\qquad\,\,+  2 \tilde{M}_{Q}^2(\eta)\frac{\gamma/\pi}{(\omega_+(Q) - \omega)^2+\gamma^2}\biggr] \ \ .
	\label{Sqwdelt}
\end{split}
\end{equation}

We are now in a position  to discuss the relative intensity of the split-off branch as compared to the primary branch. It is convenient then to define
\begin{equation}
S(Q,\omega) = S_-(Q,\omega) + S_+(Q,\omega) \ \ , 
\end{equation}
where $-(+)$ refers to the contribution to Eq.~(\ref{Sqwdelt}) of the primary (split-off) branch. We further define the maximum intensity $I_{\pm}(Q)$ of the two branches at wave vector  $Q$ as  
\begin{equation}
I_\pm(Q) = \max \{S_\pm(Q,\omega)|\,\omega > 0 \} \ \ .
\end{equation}
Choosing $\eta=0.001$ corresponding to approximately $1/20$ of the plateau width, we have plotted in Fig.[\ref{crosssection}] the structure factor at $ql_0=2.5$, which is near the minimum of the split-off dispersion, and $ql_0=1.75$, which is slightly lower than the band-crossing point at $ql_0\approx 2.1$. In the former, the relative intensity of the split-off line compared to the primary line is $\approx 15\%$ due to a large density of states of the secondary line, and the peak broadening is sufficiently small that the primary and secondary peaks can be clearly resolved. For the latter, the secondary peak intensity is $\approx 6\%$ of the primary peak. The decreased amplitude is due to the $q$ dependence of the matrix element $M_Q$, together with the absence of benefit from the van-Hove point in the density of states.

\begin{figure}[tbp]
\centering\includegraphics[width=8cm]{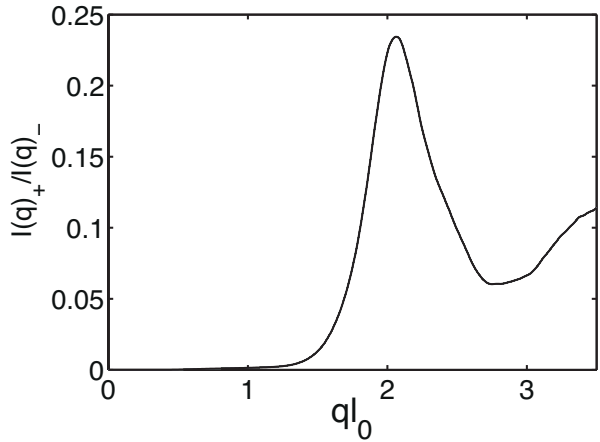}
\caption{The intensity of the split-off branch $(\omega(q)_+)$ relative to the primary branch $(\omega(q)_-)$ at filling factor $\nu=2.501$. The interplay of increased density of states near the van-Hove singularity at $ql_0\approx 2.5$, together with the increased matrix element $M_Q$ near the band crossing point at $ql_0\approx 2.1$ leads to a greatly enhanced intensity in this experimentally relevant region. For this plot, $\alpha=0.1l_0^{-1}$}
\label{intensity}
\end{figure}
In Fig.[\ref{intensity}], the relative intensity of the secondary peak compared to the primary peak is displayed over a range of experimentally relevant values of $ql_0$. The peak in relative intensity for $q \ell_0 \approx 2$ is due to the increased matrix element $\tilde{M}_Q$ near the band crossing point, together with the increased density of states for the flatter secondary branch. For small values of  $q \ell_0$, the matrix element $\tilde{M}^2_Q$ is  of order $\eta$ and hence the split-off peak has only a small visibility. When the roton mode splitting becomes comparable to and then smaller than $[t_d(\eta)^2]_{av}$ with increasing $q \ell_0$ however, the matrix element plateaus at a  value of $\tilde{M}_Q^2(\eta) = 0.0625$, corresponding   to a relative intensity of the secondary branch of $12.5\%$. 

In conclusion, we find that the 
split-off branch has a reasonably large intensity already for relatively small concentrations of qps, and at the same time the broadening of the two branches is small enough such that they can be experimentally resolved. 

\subsection{Pinning frequency of localized $e/4$ QPs}

Localized QPs are required to make  the observation of the roton split off branch possible, and one has to exclude the possibility that the collective magneto-phonon excitation of these pinned QPs could be difficult to disentangle from the 
the roton mode itself because it might occur at a comparable energy.  Here, we use the theory of the 
pinned  high magnetic field Wigner crystal \cite{Fertig99,Chitra+98,FoHu00} to 
estimate the energy of the QP pinning mode. We find that  
 it is at least one order of magnitude lower in energy than the roton minimum and hence  will not interfere with the observation of the roton minimum. We first estimate the pinning frequency $\omega_0$ of a Wigner crystal in the absence of a magnetic field, and then use the relation  $\omega_p = \omega_0^2/\omega_c$ to obtain the pinning frequency in a high magnetic field. This relation is valid if the  cyclotron frequency $\omega_c = e B / m^*$  much larger than the pinning frequency, $\omega_c \gg \omega_0$. In zero field, we have 
$\omega_0 = c_t 2 \pi /{\cal L}$,  where $c_t = \sqrt{\mu_t / n m^*}$ is the transverse phonon velocity, and 
$\mu_t = 0.245 e^2 n^{3/2}/4 \pi \epsilon_0 \epsilon$ is the shear modulus for a Wigner crystal of classical charge $e$ particles \cite{BoMa77}, $\epsilon = 12.8$ is the dielectric constant of GaAS, and ${\cal L}$ the Larkin length relevant for pinning \cite{Fertig99,Chitra+98,FoHu00}.  The effective mass of the pinned objects cancels in the final expression for $\omega_p$, such that no assumption about the effective mass of charge $e/4$ QPs is needed. 
The classical shear modulus for the Wigner crystal probabely overestimates the true shear modulus as the softening of the crystal by both quantum and thermal fluctuations is neglected. We obtain the pinning frequency $f_p = \omega_p/2 \pi$ of a high magnetic field Wigner crystal of charge $e/4$ Qps with filling fration $\eta$ as 
%
\begin{eqnarray}
f_p & = &\sqrt{3 \over 2 \pi} \ { 0.245 \over 2^8 \pi }\ {e^{2}  \over \epsilon \epsilon_0 \ell_B \hbar} \ \eta^{3/2} {1 \over {\cal L}_a^2} \nonumber \\  
&= & 3.9 \cdot  10^{10} \ {\rm Hz} \   \eta^{3/2} \sqrt{B/ 5 T} {1 \over {\cal L}_a^2} \ \ , 
\end{eqnarray}
%
where ${\cal L}_a$ is the pinning length in units of the average  QP distance $a = 2 \ell_B \sqrt{ \pi /3 \eta}$. Experimentally, it was found that the pinning mode of electron Wigner crystals is quite sharply defined \cite{Ye+02}, indicating ${\cal L}_a \gg 1$. Here, we use the conservative estimate  ${\cal L}_a = 1$ giving rise to an upper bound of the pinning frequency. For $\eta = 0.001$ used in the previous section, and $B=5 T$ we thus find $f_p <  1.2$ MHz. Even for  a QP filling fraction $\eta = 0.01$ corresponding to half the plateau width, we obtain $f_p < 39$ MHz. These frequencies are much smaller than the frequency of the roton mode obtained from $2 \pi \hbar f_{\rm roton} \approx \Delta/2$. For a gap $\Delta = 500$ mK, one finds $f_{\rm roton} 
= 5.2 $ GHz, comparable to the roton frequencies measured in \cite{smetp}, and much larger than the upper bounds for 
pinning mode frequencies.

We would like to note that the pinning mode frequencies measured for electron Wigner crystals with $\nu \lesssim 1/5$ in the lowest Landau level \cite{Ye+02}  were all in the regime of  1-3 GHz and were found to be in agreement with the above theory for values ${\cal L}_a \gg 1$. 
Further support for the fact that fractional quantum Hall pinning mode frequencies are much smaller than 
the roton frequency comes from the experiment \cite{smetp}, in which no pinning mode was observed for frequencies on the scale of the roton mode.

\end{document}